\documentstyle[12pt]{article}
\begin{document}
\large
\rightline{UT-688,1994}
\vfil
\Large
\begin{center}
Phase Operator for the Photon Field\\ and an Index Theorem
\\
\vfil
Kazuo Fujikawa
\\

Department of Physics, University of Tokyo\\
Bunkyo-ku, Tokyo 113, Japan
\vfil
Abstract

\end{center}
\large
An index relation $dim\ ker\ a^{\dagger}a - dim\ ker\ aa^{\dagger} = 1$ is
satisfied by the creation and annihilation operators $a^{\dagger}$ and
$a$
of a harmonic oscillator. A hermitian phase operator, which
inevitably leads to  $dim\ ker\ a^{\dagger}a - dim\ ker\ aa^{\dagger} = 0$,
cannot be consistently defined. If one considers an $s+1$ dimensional
truncated theory,
a hermitian phase operator of Pegg and Barnett which carries a vanishing
 index  can be  defined. However, for arbitrarily  large  $s$,
we show that
the vanishing index of the hermitian phase operator of Pegg and Barnett causes
a substantial deviation from minimum uncertainty  in a characteristically
quantum domain with small average photon  numbers.
We also mention an interesting analogy between the present
problem and the chiral anomaly in gauge theory which is related to the
Atiyah-Singer index theorem. It is suggested that the phase operator problem
related to the above analytic index may be regarded as a new class of
quantum anomaly. From an anomaly view point ,it is not surprising
that the phase operator of Susskind and Glogower, which carries a unit index,
leads to an anomalous identity and an anomalous commutator.
\\
\newpage

\section{Introduction}
\par
The quantum phase operator has been studied by various authors in the
past[1$\sim$ 10]. We here add yet  another remark on
this much studied subject, in particular, the absence of a hermitian
phase operator and the lack of a mathematical basis for
$\Delta N\Delta\phi \geq 1/2$,  on the basis of a notion of
index or an index theorem. To be specific,
we study the simplest one-dimensional harmonic oscillator defined by
\begin{eqnarray}
h&=&\frac{1}{2} (a^{\dagger}a + aa^{\dagger}) \nonumber\\
 &=&  a^{\dagger}a + 1/2
\end{eqnarray}
where $a$ and $a^{\dagger}$ stand for the annihilation and creation
operators satisfying  the standard commutator

\begin{equation}
[a, a^{\dagger} ] = 1
\end{equation}
The vacuum state $|0\rangle$
is annihilated by $a$
\begin{equation}
a|0 \rangle = 0
\end {equation}
which ensures the absence of  states with negative norm.
The number operator defined by
\begin{equation}
N = a^{\dagger}a
\end{equation}
then has non-negative integers as eigenvalues, and the annihilation
operator
$a$ is represented by
\begin{equation}
a  = |0\rangle \langle 1|+ |1\rangle \langle 2|\sqrt{2}
              + |2\rangle \langle 3|\sqrt{3} + ....
\end{equation}
in terms of the eigenstates $|k\rangle$ of the number operator
\begin{equation}
N|k\rangle = k|k\rangle
\end{equation}
with $k = 0, 1, 2, ... $. The creation operator $a^{\dagger}$ is given
by the hermitian conjugate of $a$ in (5).

 The notion of index or an index theorem provides a powerful machinery
for an
analysis of  the representation of  linear operators such as $a$
and  $a^{\dagger}$.
In the representation of $a$ and $a^{\dagger}$ specified above we have the
index condition[11]
\begin{equation}
dim\ ker\ a^{\dagger}a - dim\ ker\ aa^{\dagger} = 1
\end{equation}
where $dim\ ker\ a^{\dagger}a$ ,for example, stands for the number of
normalizable basis vectors $u_{n}$ which satisfy $a^{\dagger}au_{n}=0$
;$dim\ ker\ a^{\dagger}a$ thus agrees with the number of zero eigenvalues
of the hermitian operator $a^{\dagger}a$.
To assign a sensible meaning to the index relation, we  need a well-defined
kernel with
\begin{equation}
dim\ ker\ a^{\dagger}a  <  \infty
\end{equation}
since $dim\ ker\ a^{\dagger}a =  \infty$ corresponds to a singular point of
the index relation, which
takes place in a quantum deformation of the oscillator  algebra [12], for
example. To be mathematically precise, it is also important to confirm that
the operator $a^{\dagger}a$ has discrete eigenvalues.
In the conventional notation of index theory, the relation (7) is
written by using the trace of well-defined operators as
\begin{equation}
Tr(e^{-a^{\dagger}a/M^{2}}) - Tr(e^{-aa^{\dagger}/M^{2}}) = 1
\end{equation}
with $M^{2}$ standing for a positive constant. If $Tr(e^{-a^{\dagger}a/M^{2}})$
diverges,one needs a different regularization scheme.

 The relation(9) is confirmed for the standard representation(5) as
\begin{equation}
1 + (\sum_{n=1}^{\infty}e^{-n/M^{2}}) - (\sum_{n=1}^{\infty}e^{-n/M^{2}})= 1
\end{equation}
independently of the value of $M^{2}$.

If one should suppose the existence of a well defined hermitian phase operator
$\phi$, one would have a polar decomposition
\begin{equation}
a = U(\phi)H = e^{i\phi}H
\end{equation}
as was originally suggested by Dirac[1]. Here  $U$ and $H$
stand for unitary and hermitian operators, respectively.
To avoid the complications related to the periodicity problem, we exclusively
deal with the operator of the
form $e^{i\phi}$ in this paper; a hermitian phase operator thus means a unitary
$e^{i\phi}$, and a non-hermitian phase operator means a non-unitary
 $e^{i\phi}$ .

If (11) should be valid, one has
\begin{equation}
aa^{\dagger} = UH^{2}U^{\dagger}
\end{equation}
which is unitary equivalent to $a^{\dagger}a = H^{2}$;
$a^{\dagger}a$ and $aa^{\dagger}$ thus have an
identical number of zero eigenvalues. In this case, we have in the same
notation as (9)
\begin{eqnarray}
&&Tr(e^{-a^{\dagger}a/M^{2}}) - Tr(e^{-aa^{\dagger}/M^{2}})\nonumber\\
&=&Tr(e^{-H^{2}/M^{2}}) - Tr(e^{-UH^{2}U^{\dagger}/M^{2}})= 0
\end{eqnarray}
This relation when combined with (9) constitutes a proof of the absence of
a hermitian phase operator in the framework of index theory.

The basic utility of the notion of index or an index theorem lies
in the fact that the index as such is an integer and remains invariant
under a wide class of continuous deformation.
If one denotes two unitary equivalent operators by $a$ and $a^{\prime}$,
one has the relation
\begin{equation}
a = V^{\dagger}a^{\prime}U
\end{equation}
with $V$ and $U$ two unitary matrices. For the operator with a non-zero
index , the left and right vector spaces may in general be different,
which is the origin of the appearance of two unitary matrices $V$ and $U$
in (14). If (14) is valid, one has the relations
\begin{eqnarray}
a^{\dagger}a &=& U^{\dagger}{a^{\prime}}^{\dagger}a^{\prime}U\nonumber\\
aa^{\dagger} &=& V^{\dagger}a^{\prime}{a^{\prime}}^{\dagger}V
\end{eqnarray}
and thus $dim\ ker\ a^{\dagger}a
= dim\ ker\ {a^{\prime}}^{\dagger}a^{\prime}$
and $dim\ ker\ aa^{\dagger} = dim\ ker\ a^{\prime}{a^{\prime}}^{\dagger}$
,namely, $a$ and $a^{\prime}$ have an identical index.
As an example, the unitary time development of $a$ and $a^{\dagger}$ dictated
by the Heisenberg equation of motion, which includes a fundamental phenomenon
such as squeezing, does not
alter the index relation.
Another consequence is that  one
cannot generally  relate the representation spaces of
annihilation operators with \underline{different}
indices by a unitary transformation.

If one truncates the representation of $a$ to any finite dimension,
for example, to an $(s + 1)$ dimension with $s$ a positive
integer, one obtains
\begin{eqnarray}
dim\ ker\ a_{s}^{\dagger}a_{s} - dim\ ker\ a_{s}a_{s}^{\dagger}& = &0
\end{eqnarray}
where $a_{s}$ stands for an $s+1$ dimensional truncation of $a$.
This relation (16) is proved  by noting that  non-vanishing eigenvalues of
$ a_{s}^{\dagger}a_{s}$ and  $  a_{s}a_{s}^{\dagger}$ are in one-to-one
correspondence:
In the eigenvalue equations
\begin{equation}
a_{s}^{\dagger}a_{s}u_{n} = \lambda^{2}_{n}u_{n}
\end{equation}
one may define $v_{n} = a_{s}u_{n}/\lambda_{n}$ for $\lambda_{n} \neq 0$.
One then obtains
\begin{equation}
a_{s}a_{s}^{\dagger}v_{n} = \lambda_{n}^{2}v_{n}
\end{equation}
by multiplying $a_{s}$ to  both hand sides of eq.(17). The eigen functions
$v_{n}$ are  correctly normalized if $u_{n}$  are normalized.
For any finite dimensional matrix , one also has
\begin{equation}
Tr(a_{s}^{\dagger}a_{s}) = Tr(a_{s}a_{s}^{\dagger})
\end{equation}
These two facts combined then lead to the statement (16), namely,
$a_{s}^{\dagger}a_{s}$ and $a_{s}a_{s}^{\dagger}$ contain the same number of
zero eigenvalues.

 From the above analysis of the index condition , the polar
decomposition (11) could be consistently
defined if one truncates the dimension of the representation
space to a finite $s+1$ dimension . This is the approach adopted by Pegg
and Barnett[8] in their definition of a hermitian phase operator. See also
Ref.[9].
 Their basic
idea is then to let $s$ arbitrarily large later.
If one denotes the truncated
operators by $a_{s}$ and $a^{\dagger}_{s}$, one obtains the relation (16),
namely
\begin{eqnarray}
&&dim\ ker\ a^{\dagger}_{s}a_{s} - dim\ ker\ a_{s}a^{\dagger}_{s}\nonumber\\
&=& Tr_{(s+1)}(e^{-a^{\dagger}_{s}a_{s}/M^{2}})
-Tr_{(s+1)}(e^{- a_{s}a^{\dagger}_{s}/M^{2}})=0
\end{eqnarray}
where $Tr_{(s+1)}$ denotes an $s+1$ dimensional trace.
This index relation holds independently
of $M^{2}$ and $s$.

One technical but important point to be noted here is that the trace in (20) is
defined in an $s+1$ dimensional space, whereas the trace in (9) is defined in
an infinite
dimensional space. To directly compare these two relations, one may
extend the operator $a_{s}$ to a suitable  infinite dimensional  $A_{s}$ in
such a
way that
\begin{eqnarray}
Tr(e^{-A_{s}^{\dagger}A_{s}/M^{2}}) - Tr(e^{-A_{s}A_{s}^{\dagger}/M^{2}})= 0
\end{eqnarray}
and,for the operators to be defined in (40) and (45) in Section 2[11],
\begin{eqnarray}
 ker\ A_{s}&=& ker\ a_{s} = \{ |0\rangle\}\nonumber\\
 ker\ A_{s}^{\dagger}&=& ker\ a_{s}^{\dagger} = \{ |s\rangle\}
\end{eqnarray}
The choice of $A_{s}$ is not unique for a given $a_{s}$.
The index relation (21),which holds for arbitrary $s$,  is then regarded as a
manifestation of the invariance
of the index under a change of the deformation parameter $s$. The limit
$s \rightarrow \infty$ of the relation (21)(and also (20))is however singular
since the
kernels in (22) are ill-defined for $s \rightarrow \infty$. We thus analyze the
case of arbitrarily large but $finite$ $s$ in the main part of
this paper; this is
presumably the case in which the phase operator of Pegg and Barnett may have
some practical use. The limit $s \rightarrow \infty$ of (20) will be discussed
in Section 5 in connection with quantum anomaly.

To assign a precise meaning to this statement of
arbitrarily large $s$, we expand a given $physical$ state $|p\rangle$ as
\begin{equation}
|p\rangle = \sum_{n=0}^{\infty}p_{n}|n\rangle
\end{equation}
which is assumed to give a finite $\langle p|N^{2}|p\rangle$
\begin{equation}
\langle p|N^{2}|p\rangle = \sum_{n=0}^{\infty}n^{2}|p_{n}|^{2} = N_{p}^{2}<
\infty
\end{equation}
in addition to the normalizability of a vector in a Hilbert space
\begin{equation}
\langle p|p\rangle = \sum_{n=0}^{\infty}|p_{n}|^{2} < \infty
\end{equation}
The number $N_{p}$ in (24) characterizes a given physical system, and $s >>
N_{p}$
specifies a sufficiently large but finite $s$.

The existence of the unitary phase factor $e^{i\phi}$ of Pegg and Barnett or
its infinite dimensional extension $e^{i\Phi}$,which is associated with
$A_{s}$, critically
depends on the vanishing index in (20) and (21), which is in turn specified by
the
state $|s\rangle$ in the kernels in (22). This fact suggests that the state
$|s\rangle$,  whatever large $s$ may be,  critically influences the algebraic
relations satisfied by the
operator $e^{i\phi}$ and, consequently, the absence or presence of minimum
uncertainty states for the operator $\phi$.
In Sections 3 and 4, we shall in fact show that the state $|s\rangle$, which
characterizes the index in (20)
, causes a substantial deviation from  minimum uncertainty  for the
phase operator of Pegg and Barnett in a characteristically quantum domain
with small average photon numbers.

In Section 5, we discuss a close analogy between the present problem of the
phase operator ,which is related to the non-trivial index in (7), and
the chiral anomaly in gauge theory which is related to the Atiyah-Singer index
theorem.

\section{Quantum Phase Operators and Uncertainty Relations}
We first summarize the definitions and basic properties of two
representative "phase" operators, namely, the one due to Susskind
and Glogower[4] and the other due to Pegg and Barnett[8]. The operator
suggested by Susskind and Glogower is
\begin{equation}
e^{i\varphi} = |0\rangle \langle 1|+ |1\rangle \langle 2|
              + |2\rangle \langle 3| + ....
\end{equation}
in terms of the eigenstates $|k\rangle$ of the number operator in (6)
This phase operator is related to the operator $a$ in (5) by
$a=e^{i\varphi}N^{1/2}$.
The analogues of cosine and sine functions
are then defined by
\begin{eqnarray}
C(\varphi)&=& \frac{1}{2}(e^{i\varphi} + (e^{i\varphi})^{\dagger})\nonumber\\
S(\varphi)&=& \frac{1}{2i}(e^{i\varphi} - (e^{i\varphi})^{\dagger})
\end{eqnarray}
Note that $e^{i\varphi}$ is a symbolic notation
since $e^{i\varphi}$ is not unitary and
$\varphi$ is not defined as a hermitian operator, as is witnessed by
\begin{eqnarray}
(e^{i\varphi})^{\dagger}e^{i\varphi}& = &1 - |0\rangle\langle 0|\nonumber\\
e^{i\varphi}(e^{i\varphi})^{\dagger}& = &1
\end{eqnarray}
One can write the operator $e^{i\varphi}$ in terms of the operator $a$ in (5)
as
\begin{equation}
e^{i\varphi}=\frac{1}{\sqrt{N+1}}a
\end{equation}
As long as one considers a representation with non-negative $N$ in (29),one
obtains
the index relation[11]
\begin{eqnarray}
&&dim\ ker\ e^{i\varphi} - dim\ ker\ (e^{i\varphi})^{\dagger}\nonumber\\
&=&dim\ ker\ a - dim\ ker\ a^{\dagger}=1
\end{eqnarray}
namely, the operator $e^{i\varphi}$ carries a unit index which can be
confirmed by the explicit expression in (26).  The index relation (30)
is also written as[11]
\begin{equation}
dim\ ker\ (e^{i\varphi})^{\dagger}e^{i\varphi} - dim\ ker\ e^{i\varphi}
(e^{i\varphi})^{\dagger} = 1
\end{equation}
This last form of index relation is directly related to eq.(28),which is
in turn responsible for an anomalous commutator
\begin{equation}
{[}C(\varphi) , S(\varphi)] =  \frac{1}{2i}|0\rangle \langle 0|
\end{equation}
and an anomalous identity
\begin{equation}
C(\varphi)^{2} + S(\varphi)^{2} = 1 - \frac{1}{2}|0\rangle \langle 0|
\end{equation}
satisfied by the hermitian operators in (27). One thus sees a  relation between
the non-trivial index of $a$ in (7) and  the anomalous behaviour
of $C(\varphi)$ and $S(\varphi)$. The modified trigonometric
operators also satisfy the commutation relations with the number operator $N$,
\begin{eqnarray}
{[} N  , C(\varphi)]        & = & -i S(\varphi) \nonumber \\
{[} N  , S(\varphi)]        & = &  i C(\varphi)
\end{eqnarray}

On the other hand, the genuine hermitian phase operator $\phi$ of
Pegg and Barnett[8]
is defined in a truncated $s+1$ dimensional space by
\begin{equation}
e^{i\phi}=|0\rangle \langle 1| + |1\rangle \langle 2| + ...
          + |s-1\rangle \langle s|
          + e^{i(s+1)\phi_{0}}|s\rangle \langle 0|
\end{equation}
where $\phi_{0}$ is an arbitrary constant c-number. The operator $e^{i\phi}$
is in fact unitary in an $s+1$ dimension
\begin{equation}
e^{i\phi}(e^{i\phi})^{\dagger}=(e^{i\phi})^{\dagger}e^{i\phi}=1
\end{equation}
One may then define cosine and sine operators by
\begin{eqnarray}
\cos\phi &=& \frac{1}{2} (e^{i\phi} + e^{-i\phi})\nonumber\\
\sin\phi &=& \frac{1}{2i}(e^{i\phi} - e^{-i\phi})
\end{eqnarray}
with $e^{-i\phi} = (e^{i\phi})^{\dagger}$.
These operators together with the number operator satisfy the commutation
relations
\begin{eqnarray}
{[} N, \cos\phi] &=& -i\sin\phi\nonumber\\
&+&\frac{(s+1)}{2}[e^{i(s+1)\phi_{0}}|s\rangle \langle 0|
                 -e^{-i(s+1)\phi_{0}}|0\rangle \langle s|]\nonumber\\
{[} N, \sin\phi] &=&  i\cos\phi\nonumber\\
&-&i\frac{(s+1)}{2}[e^{i(s+1)\phi_{0}}|s\rangle \langle 0|
                 +e^{-i(s+1)\phi_{0}}|0\rangle \langle s|]\nonumber\\
{[} \cos\phi, \sin\phi] &=& 0
\end{eqnarray}
and
\begin{equation}
\cos^{2}\phi + \sin^{2}\phi = 1
\end{equation}
The $s+1$ dimensional truncated operator defined by
\begin{equation}
a_{s}=e^{i\phi}(N)^{1/2}=|0\rangle\langle 1| + |1\rangle\langle 2|\sqrt{2}
                         + .... + |s-1\rangle\langle s|\sqrt{s}
\end{equation}
and its hermitian conjugate
$a_{s}^{\dagger}$ satisfy  the algebra
\begin{equation}
[a_{s} , a_{s}^{\dagger}] = 1 - (s+1)|s\rangle \langle s|
\end{equation}
which suggests $a_{s}a_{s}^{\dagger}|s\rangle = 0$, and thus leads to the index
condition(20)
\begin{equation}
dim\ ker\ a^{\dagger}_{s}a_{s} - dim\ ker\ a_{s}a^{\dagger}_{s}=0
\end{equation}
as is required by a general analysis (13).
The operator $e^{i\phi}$ cannot be expressed in a form analogous to (29),but
the unitary operator inevitably carries a trivial index
\begin{eqnarray}
&&dim\ ker\ e^{i\phi} - dim\ ker\ (e^{i\phi})^{\dagger}\nonumber\\
&=&dim\ ker\ (e^{i\phi})^{\dagger}e^{i\phi} -
dim\ ker\ e^{i\phi}(e^{i\phi})^{\dagger}=0
\end{eqnarray}
The unitary operator simply re-labels the names of basis vectors with the
number of basis vectors kept fixed; namely,\\
 $dim\ ker\ e^{i\phi} = dim\ ker\ (e^{i\phi})^{\dagger} = 0$.
 One can confirm that the state $|s\rangle$ plays a crucial role in specifying
the indices of $a_{s}$ in (42) and $e^{i\phi}$ in
(35); the trivial index of $e^{i\phi}$ is related to the unitary property of
$e^{i\phi}$,which  in turn leads to the normal algebraic relations between
$cos\phi$ and $sin\phi$ in (38) and (39).

 From a view point of index theory, it is convenient to define an infinite
dimensional operator $e^{i\Phi}$ which has the same characteristics as
$e^{i\phi}$ in (35). One possible choice may be
\begin{eqnarray}
e^{i\Phi}&=& e^{i\phi}\nonumber\\
         &+&|s+1\rangle\langle s+2| + .....+ |2s\rangle\langle 2s+1|\nonumber\\
         &+&e^{i\phi_{1}}|2s+1\rangle\langle s+1|\nonumber\\
         &+&|2s+2\rangle\langle 2s+3| +  .... +|3s+1\rangle\langle 3s+2|
            \nonumber\\
         &+&e^{i\phi_{2}}|3s+2\rangle\langle 2s+2| +.....
\end{eqnarray}
where $e^{i\phi}$ is given in (35) and  $\phi_{1}, \phi_{2},....$ are real
constants.
This operator is unitary $e^{i\Phi}(e^{i\Phi})^{\dagger}=(e^{i\Phi})^{\dagger}
e^{i\Phi}= 1$, and $\langle k|e^{i\Phi}|k\rangle = 0$ for any $k$.The operator
$e^{i\Phi}$ gives rise to
\begin{eqnarray}
A_{s}&=&e^{i\Phi}(N)^{1/2}\nonumber\\
     &=&a_{s}\nonumber\\
     &+&|s+1\rangle\langle s+2|\sqrt{s+2}
 +...+|2s\rangle\langle 2s+1|
\sqrt{2s+1}\nonumber\\
     &+&e^{i\phi_{1}}|2s+1\rangle\langle s+1|\sqrt{s+1}\nonumber\\
     &+&|2s+2\rangle\langle 2s+3|\sqrt{2s+3} + ....
\end{eqnarray}
where $a_{s}$ stands for the operator in (40).

One can then confirm the index relation
\begin{eqnarray}
Tr(e^{-A_{s}^{\dagger}A_{s}/M^{2}}) - Tr(e^{-A_{s}A_{s}^{\dagger}/M^{2}})= 0
\end{eqnarray}
to be consistent with (13).
Apparently, $A_{s}$ is not unitary equivalent to $a$ in (5) for arbitrarily
large $s$; the limit $s \rightarrow \infty$ is however a singular point of
(46) since $ker\ A_{s}^{\dagger} = ker\ a_{s}^{\dagger} = \{|s\rangle \}$
is ill-defined in this limit. In our explicit analyses of uncertainty
relations in the next section, $e^{i\phi}$ and $e^{i\Phi}$ give rise to the
same physical consequences for sufficiently large but finite $s$ for any
physical
state which satisfies (24).

We next recapitulate a derivation of the Heisenberg uncertainty
relation for the
commutator
\begin{equation}
[A , B] = iC
\end{equation}
where $A, B$ and $C$ are hermitian operators. The expectation values
of these operators, which are real numbers, are given by
\begin{eqnarray}
\langle A\rangle &=& \langle p| A |p\rangle \nonumber\\
\langle B\rangle &=& \langle p| B |p\rangle \nonumber\\
\langle C\rangle &=& \langle p| C |p\rangle
\end{eqnarray}
for a suitable state $|p\rangle$, which is assumed to give well-defined
expectation values in (48). We then define the operators
\begin{eqnarray}
\hat{A} &=& A - \langle A\rangle\nonumber\\
\hat{B} &=& B - \langle B\rangle
\end{eqnarray}
which satisfy the same algebra as (47)
\begin{equation}
[\hat{A} ,\hat{B}] = iC
\end{equation}

We consider a function $f(t)$ defined by
\begin{equation}
f(t)= \sum_{n}|\langle n|\hat{A}|p\rangle
               - it\langle n|\hat{B}|p\rangle |^{2}
\end{equation}
where $t$ is a real parameter and the sum over $n$ runs over all the
eigenstates of the number operator. The function $f(t)$ is rewritten as
\begin{eqnarray}
f(t)&=& \sum_{n}[\langle p|\hat{A}|n\rangle \langle n|\hat{A}|p\rangle
          + t^{2}\langle p|\hat{B}|n\rangle \langle n|\hat{B}|p\rangle
                                                             \nonumber\\
        & &\ - it(\langle p|\hat{A}|n\rangle \langle n|\hat{B}|p\rangle
                -\langle p|\hat{B}|n\rangle \langle n|\hat{A}|p\rangle)]
                                                             \nonumber\\
    &=& t^{2}\langle p|\hat{B}^{2}|p\rangle
        + t\langle p|C|p\rangle + \langle p|\hat{A}^{2}|p\rangle
\end{eqnarray}
where we used the hermiticity of $\hat{A}$ and $\hat{B}$, and also the
algebra (50). By definition, $f(t)$ is positive semi-definite
\begin{equation}
f(t)\geq 0
\end{equation}
as a quadratic function of real variable t. Consequently, the discriminant
of $f(t)$ is non-negative
\begin{equation}
D= \langle p|\hat{A}^{2}|p\rangle \langle p|\hat{B}^{2}|p\rangle
 -\frac{1}{4}(\langle p|C|p\rangle)^{2} \geq 0
\end{equation}
which gives rise to the uncertainty relation
\begin{equation}
\Delta A\Delta B \geq \frac{1}{2}|\langle p|C|p\rangle|
\end{equation}
if one defines $\Delta A$ and $\Delta B$ by
\begin{eqnarray}
(\Delta A)^{2}&\equiv& \langle p|\hat{A}^{2}|p\rangle
              = \langle p|A^{2}|p\rangle - (\langle p|A|p\rangle)^{2}
                                                      \nonumber\\
(\Delta B)^{2}&\equiv& \langle p|\hat{B}^{2}|p\rangle
              = \langle p|B^{2}|p\rangle - (\langle p|B|p\rangle)^{2}
\end{eqnarray}
The necessary and sufficient condition for the equality in (55) (i.e.,
the minimum uncertainty state)[13] is
\begin{equation}
\langle n|\hat{A}|p\rangle - it\langle n|\hat{B}|p\rangle = 0
\end{equation}
for \underline{all} $n$ with a fixed real $t$ given by
\begin{equation}
t=-\frac{\langle p|C|p\rangle}{2\langle p|\hat{B}^{2}|p\rangle}
 =-\frac{2\langle p|\hat{A}^{2}|p\rangle}{\langle p|C|p\rangle}
\end{equation}

To analyze the uncertainty relation involving the number operator
$N$ ,the state $|p\rangle$ in the Hilbert space is required to satisfy the
condition (24), namely, $\langle p|N^{2}|p\rangle = \sum_{n}n^{2}|p_{n}|^{2} <
\infty$.
We call such states $|p\rangle$ as "physical states" hereafter. The
condition (24) in particular suggests
\begin{equation}
\lim_{n\rightarrow \infty}n^{3}|p_{n}|^{2} = 0
\end{equation}

\section{Minimum Uncertainty States and Index Condition}
We explain why the presence or absence of the minimum uncertainty
state can be a good characteristic of two different phase operators
with different indices. For this purpose, we start with an analysis of
 the construction of matrix elements of various operators. In particular,
we briefly explain how infinite dimensional operators are defined starting
from finite domensional ones. By this way, one can clearly understand a
special role played by the state $|s\rangle$ in the hermitian phase
operator $e^{i\phi}$.

We thus analyze  a set of matrix elements defined for
sufficiently large but finite $s$
\begin{equation}
\{ \langle n|O|p\rangle\: | n \in \Sigma_{s} \}
\end{equation}
where the operator $O$ generically stands for one of $(s+1)$ dimensional
operators,
$a_{s}$, $a_{s}^{\dagger}$ in (40) and the phase variables
$\cos\phi$ and $\sin\phi$ defined in (37).
The state $|p\rangle$ is any physical state satisfying (24): To be
precise one may have to cut-off the states in (23) at $p_{s}$, but
this does not influence our analysis below.
The  matrix elements of operators such as $a^{\dagger}a$ and
$aa^{\dagger}$ are then constructed from the matrix elements in (60).
(An analysis of more general cases than in (60) is possible, but the
analysis of (60) is sufficient for the discussion of uncertainty
relations in the present paper).
$\Sigma_{s}$ is a set of non-negative integers smaller than $s+1$
\begin{equation}
\Sigma_{s} =\{1,2, ...,s\}
\end{equation}
with $s$ standing for a cut-off parameter.

The  conventional oscillator variables and the associated phase operator
, which lead to the index relation (7),
are  specified by the set of matrix elements
\begin{equation}
\{ \langle n|O|p\rangle\: | n\in \Sigma_{s}^{\prime}\}
\end{equation}
in the limit $s\rightarrow \infty$, where $\Sigma_{s}^{\prime}$ is
\underline{any subset} of $\Sigma_{s}$ ( $\Sigma_{s}^{\prime}\neq
\Sigma_{s}$ ),
which covers all the non-negative
integers in the limit $s\rightarrow \infty$. For example,
$\Sigma_{s}^{\prime}= \Sigma_{s-1}$ or $\Sigma_{[s/2]}$ with $[s/2]$
the largest integer not exceeding $s/2$, etc. This specification of
the
operators  presumes a \underline{uniform} convergence of the
set of matrix elements (62) with respect to the choice of
$\Sigma_{s}^{\prime}$
for $s\rightarrow \infty$.
In other words, one abstracts only those properties which are independent
of the precise value of the cut-off parameter $s$ when
$s\rightarrow \infty$.
In the present case, one can confirm that the set
of matrix elements thus defined for $\cos\phi$, for example,in fact
reproduces the set of matrix elements
 of the infinite dimensional operator $C(\varphi)$
in (27) . The hermiticity of the
phase operator $\phi$ is lost in this limiting procedure and $\phi$
is converted to $\varphi$. (A related analysis from a view point of
quantum anomaly is found in Section 5).

On the other hand,the truncated space
 which leads to the representation of $a$ and $a^{\dagger}$ with
 an index $0$ to ensure the presence of a hermitian phase operator is
specified by  (60) ;one uses a \underline{very specific} $\Sigma_{s}$ for
$s\rightarrow large$ and  the uniformity of
the convergence of the set of
matrix elements is not satisfied.
The characteristic property of the operators specified by $\Sigma_{s}$, which
includes $|s\rangle$, arises from the fact that for arbitrarily large $s$
\begin{equation}
\langle s|\cos\phi|p\rangle \neq 0
\end{equation}
in general for $\cos\phi$ , and similarly for $\sin\phi$ ,in (37).
Since one cannot discriminate $|s-1\rangle$ from $|s\rangle$ in the limit
$s \rightarrow \infty$, the limit $s \rightarrow \infty$ is not well-defined
in the present case. See papers in refs.[9] and [10] for the discussions of
this problem . In the present paper we simply keep $s$ arbitrarily large but
finite for the operators specified by $\Sigma_{s}$ and  analyze their physical
implications.

  The state $|s\rangle$ is responsible for
the different indices of $e^{i\varphi}$ and $e^{i\phi}$, as was
explained in Section 2,  and also for the difference between the algebraic
relations satisfied  by $\varphi$
in (32) and (33) and the algebraic relations satisfied by $\phi$ in
(38) and (39). The absence or presence of minimum uncertainty states
is  related to those algebraic properties,
 and thus it becomes
a good characteristic of operators carrying different indices.

We next examine the minimum uncertainty state by choosing $A$ in (47) as the
number
operator
\begin{equation}
\hat{A} = \hat{N} = N - \langle p|N|p\rangle
\end{equation}
and $B$ as one of the phase operators;to be specific, we choose
$C(\varphi)$ or $\cos\phi$ in (27) and (37), respectively. The condition
for the minimum uncertainty state in (57) then becomes
\begin{equation}
\langle n|\hat{N}|p\rangle -it\langle n|\hat{C(\varphi)}|p\rangle = 0
\end{equation}
for all $n$, or
\begin{equation}
\langle n|\hat{N}|p\rangle -it^{\prime}\langle n|\hat{\cos}\phi|p\rangle = 0
\end{equation}
for all $n \leq s$. Here the
real parameters $t$ and $t^{\prime}$ are generally different. In (66)
we first fix $s$ and impose the relation for \underline{all} $n=0\sim s$ ,
 and later $s$ is set to arbitrarily large  compared to the typical number
$N_{p}$ associated with the physical state $|p\rangle$. ( If one uses an
infinite dimensional $\cos\Phi$, which is defined in terms of $e^{i\Phi}$
 (44), in (66) instead of $\cos\phi$, one can treat (65) and (66) in a
unified manner; $\cos\phi$ and $\cos\Phi$ give rise to the same physical
results for a physical state $|p\rangle$ for sufficiently large $s$ ).

One  expects that it is easier to satisfy the condition (65) than the
condition (66), since the matrix elements $\langle n|\hat{N}|p\rangle$
and $\langle n|\hat{C(\varphi)}|p\rangle$ are defined such that a uniform
convergence for
$n\rightarrow \infty$ is ensured. On the other hand,
the matrix
element $\langle n|\hat{\cos}\phi)|p\rangle$ of Pegg and Barnett
critically depends on the specific number $n=s$, whatever large $s$ may
be, while $s$ does not play a special role in $\langle n|\hat{N}|p\rangle$.

We now formulate the above qualitative consideration in a more
explicit
quantitative way. By recalling the definitions in (27) and (37), one can
confirm that the expectation values of two different phase operators
for a
 physical state are identical for sufficiently large $s$,
\begin{eqnarray}
\langle p|C(\varphi)|p\rangle &=& \langle p|\cos\phi|p\rangle \nonumber\\
\langle p|S(\varphi)|p\rangle &=& \langle p|\sin\phi|p\rangle
\end{eqnarray}
The variation of the number operator $N$ is of course common to
both cases for sufficiently large $s$, and it is given by
\begin{equation}
(\Delta N)^{2} =\langle p|N^{2}|p\rangle - (\langle p|N|p\rangle)^{2}
\end{equation}
However, the expectation values of the square of phase variables are
generally different
\begin{eqnarray}
\langle p|C(\varphi)^{2}|p \rangle
&=& \sum_{n=0}^{n_{p}}\langle p|C(\varphi)|n\rangle
                      \langle n|C(\varphi)|p\rangle \nonumber\\
\langle p|\cos^{2}\phi|p \rangle
&=& \sum_{n=0}^{n_{p}}\langle p|cos\phi|n\rangle
                      \langle n|cos\phi|p\rangle \nonumber\\
& &                  +\langle p|cos\phi|s\rangle
                      \langle s|cos\phi|p\rangle \nonumber\\
&=& \sum_{n=0}^{n_{p}}\langle p|C(\varphi)|n\rangle
                      \langle n|C(\varphi)|p\rangle \nonumber\\
& &                  +|\langle p|\cos\phi|s\rangle|^{2}\nonumber\\
&\geq& \langle p|C(\varphi)^{2}|p\rangle
\end{eqnarray}
where $n_{p}$ stands for the maximum occupation number which contributes
to the matrix element of $C(\varphi)$ sizably for a given physical state
$|p\rangle$: One may choose $n_{p}>> N_{p}$ in (24). We also used the fact that
the operator $\cos\phi$ is
hermitian and that $\langle n|\cos\phi|p\rangle
= \langle n|C(\varphi)|p\rangle $ for $n\leq n_{p}$ and sufficiently large $s$
{}.
 Note that $\langle s|\cos\phi |p\rangle$ is not zero in general
even for arbitrarily large $s$ due to the definition in (35).

We thus conclude from (67) and (69) that
\begin{eqnarray}
(\Delta \cos\phi)^{2}&=& (\Delta C(\varphi))^{2} +
                        |\langle p|\cos\phi|s\rangle|^{2}\nonumber\\
                     &\geq& (\Delta C(\varphi))^{2}
\end{eqnarray}
and
\begin{eqnarray}
\Delta N \Delta \cos\phi &\geq& \Delta N \Delta C(\varphi)\nonumber\\
             &\geq& \frac{1}{2}|\langle p|S(\varphi)|p\rangle|\nonumber\\
             &=&   \frac{1}{2}|\langle p|\sin\phi|p\rangle|
\end{eqnarray}
where we used the fact that
\begin{equation}
\langle p|(\frac{s+1}{2})\{e^{i(s+1)\phi_{0}}|s\rangle\langle 0|
                      -e^{-i(s+1)\phi_{0}}|0\rangle\langle s|\}|p\rangle =0
\end{equation}
in (38) for sufficiently large $s$ by noting (59).

Similarly, one can derive from (32), (34) and (38)
\begin{eqnarray}
\Delta N \Delta \sin\phi &\geq& \Delta N \Delta S(\varphi)\nonumber\\
             &\geq& \frac{1}{2}|\langle p|C(\varphi)|p\rangle|\nonumber\\
             &=&   \frac{1}{2}|\langle p|\cos\phi|p\rangle|\\
\Delta \cos\phi \Delta \sin\phi &\geq& \Delta C(\varphi) \Delta S(\varphi)
                                                         \nonumber\\
         &\geq& \frac{1}{4}|\langle p|0\rangle \langle 0|p\rangle|
                                                         \nonumber\\
         &\geq& 0
\end{eqnarray}

{}From these relations we learn that the uncertainty relations are
always better satisfied for the phase varibles $C(\varphi)$ and
$S(\varphi)$. If the state $|p\rangle$ is a minimum uncertainty state
for the variables $(\cos\phi, N)$ or $(\sin\phi,N)$ ,the same state
automatically becomes a minimum uncertainty state for the variables
$(C(\varphi),N)$ or $(S(\varphi),N)$,respectively. But the other way
around is not true in general. Also, Eq.(74) shows that  uncertainty in
$\Delta \cos\phi\Delta \sin\phi$ is always greater than or equal to
 uncertainty in
$\Delta C(\varphi)\Delta S(\varphi)$ for \underline{any} given  physical
state $|p\rangle$ despite of $[\cos\phi, \sin\phi] = 0$ in (38).
As far as the measurements of physical matrix elements are concerned,
the non-commuting property of $C(\varphi)$ and $S(\varphi)$ provides
a constraint less stringent than the commuting property of $\cos\phi$
and $\sin\phi$ which comes in pairs with $\cos^{2}\phi +\sin^{2}\phi =
1$. As for the minimum uncertainty states,
it has been shown explicitly by Jackiw[5] that a minimum
uncertainty state exists(under certain conditions) for the pair of
variables $(N,C(\varphi))$
or $(N,S(\varphi))$, but there is
no normalizable state which satisfies the minimum uncertainty relation
( in the strict sense) for the pair $(C(\varphi),S(\varphi))$ in (74)[6].

The uncertainty relations for the hermitian variable $\phi$ of Pegg
and Barnett substantially deviate from the minimum uncertainty
when the physical state $|p\rangle$ has
a substantial overlap with the vacuum state $|0\rangle$, since in this
case $\langle p|\cos\phi|s\rangle$ or $\langle p|\sin\phi|s\rangle$
becomes appreciable in (70) or in a corresponding relation. In a
characteristically quantum domain with
small particle numbers(or we use the term "photon numbers" hereafter),
 we generally have no minimum uncertainty state
for the phase operator of Pegg and Barnett except for the obvious
case $\Delta N=0$.

\section{Minimum Uncertainty States in Characteristically Quantum
Domain}

To explicitly illustrate the absence of minimum uncertainty states
for the phase variable $\phi$
in the characteristically quantum domain , we
consider the case where the average
photon number  is almost zero and the probability of having one
photon is very small with negligible probability for more than one photons.
The most
general state for this situation is given by
\begin{equation}
|\alpha\rangle \simeq \frac{1}{\sqrt{1+|\alpha|^{2}}}
                      [|0\rangle + e^{i\theta}|\alpha||1\rangle ]
\end{equation}
with $|\alpha| \ll 1$, which incidentally corresponds to a small $|\alpha|$
limit of the coherent state
\begin{equation}
|\alpha\rangle \equiv e^{-\frac{1}{2}|\alpha|^{2}}
\sum_{n=0}^{\infty}\frac{e^{in\theta}|\alpha|^{n}}{\sqrt{n!}}|n\rangle
\end{equation}
For the state $|\alpha\rangle$ in (75), one has
\begin{eqnarray}
\langle\alpha|N|\alpha\rangle &=& \frac{|\alpha|^{2}}{1+|\alpha|^{2}}
                                                      \nonumber\\
\langle\alpha|N^{2}|\alpha\rangle &=& \frac{|\alpha|^{2}}{1+|\alpha|^{2}}
                                                      \nonumber\\
\langle\alpha|C(\varphi)|\alpha\rangle &=&
                                      \langle\alpha|\cos\phi|\alpha\rangle
                                  =\frac{|\alpha|}{1+|\alpha|^{2}}\cos\theta
                                                      \nonumber\\
                                 &\simeq&|\alpha|\cos\theta\nonumber\\
\langle\alpha|S(\varphi)|\alpha\rangle &=&
                                      \langle\alpha|\sin\phi|\alpha\rangle
                                  =\frac{|\alpha|}{1+|\alpha|^{2}}\sin\theta
                                                      \nonumber\\
                                 &\simeq&|\alpha|\sin\theta\nonumber\\
\langle\alpha|C(\varphi)^{2}|\alpha\rangle&\simeq& \frac{1}{4}\nonumber\\
\langle\alpha|\cos^{2}\phi|\alpha\rangle&=& \frac{1}{2}
\end{eqnarray}
Thus
\begin{eqnarray}
\Delta N&=& \sqrt{\langle\alpha|N^{2}|\alpha\rangle -
                                      {\langle\alpha|N|\alpha\rangle}^{2}}
                                                      \nonumber\\
        &\simeq& |\alpha|                             \nonumber\\
\Delta C(\varphi)& =& \sqrt{\frac{1}{4} -
                               {\langle\alpha|C(\varphi)|\alpha\rangle}^{2}}
                                                      \nonumber\\
                 &\simeq& \frac{1}{2}                 \nonumber\\
\Delta \cos\phi &=& \sqrt{\frac{1}{2}-
\langle\alpha|\cos\phi|\alpha\rangle^{2}}
                                                      \nonumber\\
                &\simeq& \sqrt{\frac{1}{2}}
\end{eqnarray}
One therefore obtains
\begin{eqnarray}
\Delta N\Delta C(\phi)&\simeq& \frac{1}{2}|\alpha| \nonumber\\
\Delta N\Delta\cos\phi&\simeq& \sqrt{\frac{1}{2}}|\alpha|
\end{eqnarray}
and uncertainty relations by noting (72)
\begin{eqnarray}
\Delta N\Delta\cos\phi=\sqrt{\frac{1}{2}}|\alpha| > \Delta N\Delta C(\varphi)
=\frac{1}{2}|\alpha|
                      &\geq&\frac{1}{2}|\alpha||\sin\theta|\nonumber\\
& &
\end{eqnarray}
with $|\alpha||\sin\theta| = |\langle\alpha|S(\varphi)|\alpha\rangle|
= |\langle\alpha|\sin\phi|\alpha\rangle|$. The minimum uncertainty is
achieved for $|\sin\theta|=1$ for the set
of variables $(N, C(\varphi),S(\varphi))$, but no minimum uncertainty
state for the set of variables $(N, \cos\phi, \sin\phi)$ of Pegg and
Barnett.

Similarly one obtains
\begin{eqnarray}
\Delta N\Delta\sin\phi=\sqrt{\frac{1}{2}}|\alpha| &>& \Delta N\Delta S(\varphi)
=\frac{1}{2}|\alpha|\geq\frac{1}{2}|\alpha||\cos\theta|\nonumber\\
\Delta\cos\phi\Delta\sin\phi =\frac{1}{2} &>& \Delta C(\varphi)\Delta
S(\varphi) =\frac{1}{4} = \frac{1}{4}|\langle\alpha|0\rangle|^{2}
\nonumber\\
& &
\end{eqnarray}
Note that the uncertainty product $\Delta\cos\phi\Delta\sin\phi=1/2$
is essentially a consequence of $\cos^{2}\phi + \sin^{2}\phi =1$.
The measurement of the uncertainty product in the second equation of (81)
may provide a direct experimental test of the choice of a physical
phase operator.

Physically, a marked deviation from  minimum uncertainty for the
variable $\phi$ may be understood as follows: To maintain the hermiticity
of $\phi$ and make $\phi$ rotate over full $2\pi$ angle in a
characteristically quantum domain, the transition from $|0\rangle$ to
$|s\rangle$ is required as is seen from the last term in (35). This is
not possible as a real physical process for large $s$. On the other hand,
all the states up to $|s\rangle$ contribute to the intermediate states
of algebraic relations such as (38) and (39) without any suppression
factor. These two properties combined cause a severe discrepancy
between the physical matrix elements and the formal algebraic relations,
as is seen in (80) and (81).In passing, we note that the transition from
$|s+1\rangle$ to $|2s+1\rangle$, for example, in $e^{i\Phi}$ (44) is
negligible for a physical state which satisfies (59), and thus the two
hermitian phase operators $e^{i\phi}$ and $e^{i\Phi}$ give the same physical
results in this section.

This absence of the minimum uncertainty state for the phase
variable of Pegg and Barnett is shown more generally on the basis of
relations (70) and (71). A necessary condition for the minimum
uncertainty for the variable of
Pegg and Barnett is that the physical states $|p\rangle$ satisfy
\begin{equation}
\langle s|\cos\phi|p\rangle =0\ or\ \langle s|\sin\phi|p\rangle =0
\end{equation}
namely,  $|p\rangle$ do not contain the zero photon state
$|0\rangle$, or the states $|p\rangle$  spread over many
eigenstates of the number operator such that the term
$|\langle s|\cos\phi|p\rangle|^{2}$ is negligible compared with
$\sum_{n=0}^{n_{p}}|\langle n|\cos\phi|p
\rangle|^{2}$ in (69). This latter possibility is realized, for example,
by the coherent state with large $|\alpha|$.[8]

In passing, the algebraic consistency is improved for the variable of
Pegg and Barnett if one sets the cut-off parameter $s$ in the region
of average photon number. If one sets $s=1$ , for example, one obtains
\begin{eqnarray}
\langle\alpha|\cos\phi|\alpha\rangle &=&
            2|\alpha|{\cos\phi_{0}}\cos(\theta-\phi_{0})\nonumber\\
\langle\alpha|\cos^{2}\phi|\alpha\rangle &\simeq& \cos^{2}\phi_{0}
\end{eqnarray}
and, consequently, from (38) with $s=1$
\begin{eqnarray}
\Delta N\Delta\cos\phi &\simeq& |\alpha||\cos\phi_{0}| \nonumber\\
                       &\geq& \frac{1}{2}|\langle\alpha|\{
                       \sin\phi + ie^{2i\phi_{0}}|1\rangle\langle 0|-
                      ie^{-2i\phi_{0}}|0\rangle\langle 1|\}|\alpha\rangle|
                       \nonumber\\
                       &=& |\alpha||\cos\phi_{0}\sin(\theta-\phi_{0})|
\end{eqnarray}
The minimum uncertainty is achieved if
\begin{equation}
|\sin(\theta-\phi_{0})| = 1
\end{equation}
By recalling the $(s+1)$ orthonormal eigenstates of the phase variable
$\phi$[8]
\begin{equation}
|\phi_{m}\rangle = (s+1)^{-1/2}\sum_{n=0}^{s} e^{in\phi_{m}}|n\rangle
\end{equation}
with $\phi_{m}=\phi_{0} + 2\pi m/(s+1)$, $m=0\sim s$, one obtains
for $s=1$
\begin{eqnarray}
P(\phi_{0}) = |\langle\phi_{0}|\alpha\rangle|^{2}&=&
            \frac{1}{2}[1+ 2|\alpha|\cos(\theta-\phi_{0})]\nonumber\\
P(\phi_{1}) = |\langle\phi_{1}|\alpha\rangle|^{2}&=&
            \frac{1}{2}[1- 2|\alpha|\cos(\theta-\phi_{0})]
\end{eqnarray}
For the choice of $\theta-\phi_{0}$ in (85), one has a uniform phase
distribution
\begin{equation}
P(\phi_{0}) =P(\phi_{1}) =1/2
\end{equation}
This exercise shows that the algebraic consistency is improved for
$s\sim 1+|\alpha|+ |\alpha|^{2}$, but one is apparently dealing with a theory
different from the original one in (5).

\section{Analogy with Chiral Anomaly}
\par
The absence  of the
minimum uncertainty state for the operators of Pegg and Barnett in a
characteristically quantum domain arises
from their very definition and the index mismatch. This fact
may not prohibit
the use of the phase variable of Pegg and Barnett as an
interpolating variable in practical analyses for finite $s$, but it at least
shows that we cannot attach much physical
significance to the deviation from minimum uncertainty in a
characteristically quantum domain with small photon
numbers.  Our consideration shows that the notion of index
or an  index theorem provides a powerful machinery
in the analysis of the representation of linear operators.

We here comment
on an interesting analogy between the present problem and the chiral
anomaly[14][15]
 in gauge theory which is related to the Atiyah-Singer index
theorem.

In gauge theory one deals with a (Euclidean) Dirac operator
defined by
\begin{equation}
/\!\!\!\!D = \sum_{\mu=1}^{4}\gamma^{\mu}(\frac{\partial}{\partial x^{\mu}}
                                       - igA_{\mu}^{a}
                      T^{a})
\end{equation}
where $\gamma^{\mu}$ is the $4\times 4$ anti-hermitian Dirac matrix with
$\gamma_{5}\equiv \gamma^{4}\gamma^{1}\gamma^{2}\gamma^{3}$,
 $A_{\mu}^{a}(x)$
is the non-Abelian Yang-Mills field ,$g$ is the coupling constant
and $T^{a}$ is the hermitian generator of the
gauge group. The non-zero index relation for $/\!\!\!\!D_{R} =
/\!\!\!\!D (\frac{1+\gamma_{5}}{2})$ [16]
\begin{equation}
dim\ ker\ /\!\!\!\!D_{R} - dim\ ker\ /\!\!\!\!D^{\dagger}_{R} =\nu \neq 0
\end{equation}
with $\nu$ the Pontryagin index,which is expressed in terms of the gauge
field $A_{\mu}^{a}(x)$, is used as an argument for the
presence of the  chiral anomaly: The Hilbert space
for a single fermion inside the background gauge field with $\nu \neq 0$
cannot be unitary equivalent
to the Hilbert space of a free fermion with $\nu=0$.
Here the left-hand side of (90) is specified by the difference in the number
of
eigenstates of $\gamma_{5}$
\begin{equation}
\gamma_{5}\varphi_{n}(x) = \pm \varphi_{n}(x)
\end{equation}
for $\lambda_{n}=0$ in the eigenvalue problem in 4-dimensional
space-time
\begin{equation}
/\!\!\!\!D\varphi_{n}(x) = \lambda_{n}\varphi_{n}(x),\,\,\int
\varphi_{n}^{\dagger}(x)\varphi_{m}(x)\,d^{4}x = \delta_{n,m}
\end{equation}
The interaction
picture assumes the unitary equivalence of these two Hilbert spaces with
$\nu \ne 0$ and $\nu = 0$, and
thus interaction picture perturbation theory inevitably encounters a
surprise (i.e.,anomaly).

In the actual analysis of the chiral anomaly,
it is known[15][17] that a
careful treatment of the ultraviolet cut-off, which is analogous to
the parameter $s$ in the present phase operator, is required to
recognize the consequence of the index relation (90): The decoupling
or the failure of decoupling of the ultraviolet cut-off from physical
quantities needs to be analyzed with great care.
The connection
between  the non-zero index and the chiral anomaly appears in a
particularly
transparent way in the Euclidean path integral formulation of anomalies,
which is based on the analysis of single fermion states in a
background gauge field [17]. As for physics aspects, the chiral anomaly,
which is related to  high energy behavior in the interaction picture,
critically influences the low energy phenomena such as the radiative
decay of the neutral pion $\pi^{0} \rightarrow \gamma\gamma$ [15].

If one uses an analogy between the phase operator and the chiral
anomaly, the index relation (7) corresponds to the presence
of a quantum
anomaly and the relation (13) or (20) to the normal situation naively expected
by a classical consideration. The anomaly specified by the index relation
(7) is clearly recognized only when one carefully analyzes the dependence
 of the matrix elements of various operators on the cut-off parameter $s$:
For example,a sequence of the sets of matrix elements is not
uniformly convergent in the formulation of Pegg and Barnett, as is
seen in (60).

The normal situation (i.e., hermitian phase operator) realized by a finite
dimensional formulation of Pegg and Barnett in (20) may then be regarded as
corresponding to the case of
chiral anomaly where the mass of the Pauli-Villars regulator is kept
finite; the finite regulator mass generally
avoids anomalous behavior but leads to a
different theory. Also the effect of the regulator does  not quite
decouple even in the limit of arbitrarily large regulator mass (or $s$),
 which is
the origin of the discrepancy between the algebraic relations (32) and (38).

To be more explicit, one may rewrite the relation (20) as
\begin{eqnarray}
& &Tr_{(s+1)}(e^{-a_{s}^{\dagger}a_{s}/M^{2}})-
Tr_{(s)}(e^{-a_{s}a_{s}^{\dagger}/M^{2}})\nonumber\\
&=&Tr_{(s+1)}(e^{-a_{s}a_{s}^{\dagger}/M^{2}})-
   Tr_{(s)}(e^{-a_{s}a_{s}^{\dagger}/M^{2}})= 1
\end{eqnarray}
where $Tr_{(s)}$ stands for the trace over the (first) $s$-dimensional
subspace of the $(s+1)$-dimensional space; the right-hand side of (93)
is the contribution of the state $|s\rangle$.
 This relation (93) holds for
any positive $M^{2}$ and $s$, and it can be confirmed that each term in
the left-hand side has a well-defined limit for $s \rightarrow \infty$,
and one recovers (9) in this limit
\begin{equation}
Tr(e^{-a^{\dagger}a/M^{2}}) - Tr(e^{-aa^{\dagger}/M^{2}}) = 1
\end{equation}
This phenomenon is a precise analogue of the chiral anomaly; the effect
of the cut-off, i.e.,the state $|s\rangle$, gives rise to the ``anomaly''
in the right-hand side of (94) in the limit  $s \rightarrow \infty$, though
one obtains a normal relation (20) for a finite $s$.

The physical implications of the phase operator anomaly appear most
prominently in the "low energy" processes with small average photon
numbers, such as in (80) and (81); apparently ``anomalous'' behavior exhibited
by
the phase operator which carries a unit index is
in fact more consistent with quantum phenomena.
In terms of operator language,the  relation in (32) may be regarded
as an anomalous
commutator and the relation (33) as an anomalous identity: Namely we have
\begin{eqnarray}
{[}C(\varphi) , S(\varphi)]&=&  \frac{1}{2i}|0\rangle \langle 0|\nonumber\\
C(\varphi)^{2} + S(\varphi)^{2}&=& 1 - \frac{1}{2}|0\rangle \langle 0|
\end{eqnarray}
both of which
are characteristic properties of any quantum anomaly[15][18]. We also
emphasized that these properties are directly related to the non-vanishing
index in (31).
One may thus regard the phase operator problem associated with the
non-vanishing analytic index in (7) as a new class of quantum anomaly.
 From this view point , the anomalous behavior seen in (32) and (33)
is an \underline{inevitable} real quantum effect, not an artifact of
our insufficient definition of phase operator. This may be tested by
experiments by measuring the uncertainty product in (81) or its variants,
since the
phase operator of Susskind and Glogower, though quite an attractive choice,  is
one of those operators
which carry a natural index, i.e.,a unit index.

We also mention that an analogue of anomaly in the present phase operator
is based on an
analysis of multi-photon states with a fixed momentum and polarization
and thus it is characteristic to bosonic systems, whereas the conventional
anomaly in field theory is primarily based on an analysis of one-
particle states [17] which is applicable to both of fermions and
bosons. The phase operator anomaly is a qualitatively new
quantum anomaly.

It is interesting that the first paper on quantum field theory [1] already
contained an analytic index and quantum anomaly.


\begin{thebibliography}{1}
\bibitem{11}
The index relation is generally formulated as\\
$dim\ ker\ a - dim\ ker\ a^{\dagger} = 1 $ \\
instead of (7), or\\
$dim\ ker\ a_{s} - dim\ ker\ a_{s}^{\dagger} = 0$\\
for the operator $a_{s}$ in (20);$dim\ ker\ a$, for example, stands for the
number of normalizable states which are annihilated by the operator $a$. In the
body of the present paper  we use the relations (7) and (20), since they have a
more
transparent physical interpretation as a difference in the number
of zero eigenvalues of hermitian operators $a^{\dagger}a$ and
$aa^{\dagger}$:The equivalence of these two
 specifications is seen by
noting that $au = 0$ implies $a^{\dagger}au = 0$. Conversely,  $a^{\dagger}au =
0$
implies  $(a^{\dagger}au,u) = (au, au)
= 0$ and thus  $au = 0$ if the inner product is positive
definite.\\
\bibitem{13}
The minimum
uncertainty state in this paper is defined as the state which
satisfies the equality
in (55). This does not necessarily mean that $\Delta A\Delta B$ is
actually minimum for this state since the right-hand side of (55)
depends on the chosen state. The minimum uncertainty relation here is
regarded as a test of algebraic consistency of the commutator in (47)
for the set of suitable physical states.
\bibitem{14}
Quantum anomaly may be broadly understood as an
inevitable and unavoidable property of quantized operators
which contradicts  expectations on the basis of  classical-
quantum analogy.
\end{thebibliography}
\end{document}